\documentclass[aps,showpacs,a4paper,floatfix,twocolumn,prc,amsmath,amssymb]{revtex4-1}
\usepackage{graphicx}
\usepackage{epsfig}
\usepackage{ulem}
\usepackage{morefloats}
\usepackage{rotating}
\usepackage{color}
\usepackage{subfigure,dcolumn}
\usepackage{epstopdf}
\usepackage{mhchem}
\usepackage{upgreek}
\usepackage{color}
\usepackage{soul}

\def\beq{\begin{equation}}
\def\eeq{\end{equation}}
\def\beqa{\begin{eqnarray}}
\def\eeqa{\end{eqnarray}}
\def\ban{\begin{eqnarray*}}
\def\ean{\end{eqnarray*}}
\def\bi{\begin{itemize}}
\def\ei{\end{itemize}}

\newcommand{\pc}[1]{\ensuremath{\left(#1\right)}}

\begin{document}

\title{Full distribution of clusters with universal couplings and in-medium effects}
\author{Helena Pais$^1$}
\author{Francesca Gulminelli$^2$}
\author{Constan\c ca Provid{\^e}ncia$^1$}
\author{Gerd R\"opke$^{3,4}$}

\affiliation{$^1$CFisUC, Department of Physics, University of Coimbra,
  3004-516 Coimbra, Portugal. \\
$^2$LPC (CNRS/ENSICAEN/Universit\'e de Caen Normandie), UMR6534, 14050 Caen c\'edex, France. \\
$^3$Institut f\"ur Physik, Universit\"at Rostock, D-18051 Rostock, Germany. \\
$^4$National Research Nuclear University (MEPhI), 115409 Moscow, Russia.}

\begin{abstract}
Light and heavy clusters are calculated for asymmetric warm nuclear matter in a relativistic mean-field approach. In-medium effects, introduced via a universal cluster-meson coupling, and a binding energy shift contribution, calculated in a Thomas-Fermi approximation, were taken into account. This work considers, besides the standard lightest bound clusters $^4$He, $^3$He, $^3$H, and $^2$H, also stable and unstable clusters with higher number of nucleons, in the range $5\leq A \leq 12$, as it is natural that heavier clusters also form in core-collapse supernova matter, before the pasta phases set in. We show that  these extra degrees of freedom contribute with non-negligible mass fractions to the composition of nuclear matter, and  may prevail over deuterons and $\alpha$ particles at high density in strongly asymmetric matter, and not too high temperatures. The presence of the light clusters reduces the contribution of heavy clusters to a much smaller density range, and to a smaller mass fraction.

\end{abstract}


\maketitle

\section{Introduction}

Light \cite{Roepke,Raduta,Ferreira12,providencia12,virial,Hempel,typel10,pais-typel} and heavy \cite{Ravenhall1983,Maruyama2005,Watanabe2005,heavy} clusters exist in nature in different scenarios: the inner crust of neutrons stars \cite{ns}, i.e., cold $\beta-$equilibrium stellar matter, and also in warm  nuclear matter with fixed proton fraction, like core-collapse supernova (CCSN) matter \cite{oertel},  and neutron star mergers \cite{GW}. Light clusters might have an effect on the average energy of both neutrinos and antineutrinos, emitted during the supernova explosion, as they may increase or decrease it, having thus consequences on the cooling of the protoneutron star \cite{arcones08,cooling}. Actually, in Ref.~\cite{arcones08}, it was found that these clusters are the major source of opacity for antineutrinos. 
Consequently, transport properties can also be modified by these inhomogeneities, and studies with these clusters should be performed. In fact, studies \cite{arcones08} show that the outer layers of a protoneutron star have the ideal conditions for the formation of light clusters, especially tritons and deuterons, and these clusters  are not always taken into account in the equation of state (EoS) for core-collapse supernova simulations. Another site where these clusters can form are heavy-ion collisions (HIC) experiments. These sites can have similar temperatures and densities as in CCSN matter, but the asymmetry and charge content can be quite different, as it was pointed out in Ref.~\cite{hempel15}. Therefore, these terrestrial experiments can also be used to set constrains on warm non-homogeneous matter. Currently, we have only one unique existing constraint on in-medium modifications of light particle yields at high temperature, the so called chemical equilibrium constants, which was proposed from data coming from HIC \cite{qin12}. 

Recently, a new approach for the in-medium effects of these light clusters was considered \cite{pais18}. This approach is based on the single nucleus approximation, but the inclusion of light clusters is a step to go beyond the first models proposed within a similar framework \cite{Lattimer,HShen,shen2010}. In our calculation, in-medium effects are taken into account in the binding energy of the light clusters and treated self-consistently, no excluded volume is introduced, the temperature- and isospin-dependent surface energy  is consistently calculated, and the crust-core transition is appropriately described at variance with many statistical models. Besides, the present approach is easily developed in the framework of any RMF model that may be considered more adequate to reproduce nuclear matter and nuclei properties.

A  completely different approach is taken with nuclear statistical equilibrium (NSE) models which consider a full nuclear distribution in thermal equilibrium  \cite{smn,smsm,Hempel,fyss, gr2018,shen2011,schneider2017,buyukcizmeci2013}. Some more recent models already include interactions between unbound nucleons and nuclei which allow the description of matter close to the crust-core transition \cite{gr2018,Hempel,fyss}. The interaction between nucleons is undertaken within an effective nuclear interaction, and between nucleons and nuclei within the excluded volume approximation. Recently, in Ref. \cite{buyukcizmeci2013}, three NSE models have been compared for densities below $0.1\rho_0$, and it was shown that although the main trends were described by all of them, important differences did arise at the higher densities and lower temperatures, due to the different modeling of medium effects, as, for instance, the temperature and density dependence of the surface and bulk energies of heavy nuclei.
 
 Overall, we expect that our approach  will  allow the improvement of extended NSE models to  correctly include in-medium effects and  describe core-collapse supernova matter. The single-nucleus approximation with the improved treatment of few-nucleon correlations, the inclusion of interacting light clusters, is a first step in this direction.

In our model, the in-medium effects were taken into account by a modification of the scalar cluster-meson coupling, and by including an extra-term in the binding energy of the clusters (see Ref. \cite{pais18}), derived in the Thomas-Fermi approximation, which works as an excluded-volume effect. At very low densities, there is one equation of state, the Virial EoS (VEoS) \cite{virial, typel12-v}, that can be used as a model-independent constraint, since it is only based on experimentally measured scattering phase shifts and binding energies, and gives the correct zero-density limit for the EoS at finite temperature. With the increase of the density, the interactions between particles become stronger, until the VEoS is no longer valid, and, in this regime, it is the binding energy shift that plays an important role, as we showed in Ref.~\cite{pais18}.
Our model not only reproduces the Virial EoS in the low-density limit, but also the equilibrium constants extracted from experimental data coming from heavy-ion collisions \cite{qin12}. There, the following light clusters, $^4$He, $^3$He, $^3$H, and $^2$H, which were also considered in many other different studies \cite{Roepke,Raduta,Ferreira12,providencia12,virial,Hempel,typel10}, were taken into account. In the following, we will refer to these four particle species as to the ``classical" clusters.

In all the possible scenarios where the clusters may exist, by increasing the density, heavier light clusters, like $^5$He and $^5$H, are also expected to form, before the clusters become so heavy, that the pasta phases develop. 
In the present work, we want to investigate if the heavier light clusters should be considered in studies of stellar matter. To this aim, besides considering the four ``classical" light clusters, we include in our model all light clusters with $A\leq 12$. We will refer to these extra particle species as to the ``exotic" clusters. Whereas light clusters have been extensively investigated, almost nothing is done with respect to the ``exotic" light clusters. Because they are mostly weakly bound and show cluster structures (e.g. $^8$Be), it is assumed that they are strongly influenced (and suppressed) by the medium, for a discussion see \cite{Roepke85,Roepke13}. Nevertheless, they have to be discussed and this is the aim of this work.
Many of these clusters are unstable towards particle emission in the laboratory. This is not expected to influence their abundance in stellar matter, because the strong interactions are in equilibrium in the stellar medium, and so are the weak decays in the final stage of the collapse, when the density is sufficiently high for these clusters to be produced. However, when comparing to experimental data from HIC, where this equilibrium is not achieved, we take into account their decay modes into $\alpha$ and triton clusters. 

Finally, the effect of a heavy cluster (pasta) is also included within a compressible liquid drop (CLD) approach \cite{pais15}. The CLD calculation follows the same principles as the coexistence phase (CP) approximation, where the  Gibbs equilibrium conditions are imposed in order to obtain the minimum-energy state, but it considers the surface and Coulomb terms in the free energy before the minimization is done. 
A similar calculation was already introduced in a previous work \cite{avancini17}, where the authors proposed a Thomas-Fermi calculation with light clusters.

This paper is organized in the following way: in Section \ref{sec:Formalism}, the two calculations, homogeneous matter and CLD with light clusters, used throughout the work are presented. Then, in Section \ref{sec:Exotic}, we introduce the exotic clusters in our formalism, and in Section \ref{sec:A12}, we discuss the role of exotic light clusters with $4 < A \leq 12$. There, we focus on understanding how their inclusion affects the calculation of warm stellar matter, by determining  how the mass fraction of the four light clusters, $^4$He, $^3$He, $^3$H, and $^2$H is changed, and by analyzing which ones of these exotic light clusters are the most abundant. We will take into consideration the fact that many of these exotic clusters are unstable, and we evaluate their effective mass distributions by taking into account the decay rates. Lastly, in Section \ref{sec:CLD}, we show the CLD results with the inclusion of all the light clusters mentioned in this work, and, finally, in Section \ref{sec:Conclusions}, some conclusions are drawn.

\section{Formalism}\label{sec:Formalism}

In this paper, we consider two different calculations: homogeneous matter with the inclusion of light clusters \cite{Ferreira12}, and the compressible liquid drop model calculation
\cite{pais15}, where we also include light clusters. 

\subsection{Homogeneous matter with light clusters}

In our system, we consider three meson fields, the isoscalar-scalar
$\phi$, the isoscalar-vector field $\omega^{\mu}$ and the isovector-vector
$\mathbf{b}^{\mu}$, that interact with the nucleons and light cluster, both bosons and fermions, with mass number $2\leq A \leq 12$. Since we are
dealing with stellar matter, electrons must also be included to
achieve neutrality.

 The total Lagrangian density for a system that includes only the
 ``classical" clusters, deuteron ($d$), triton ($t$), helion ($h$) and
 $\alpha$,  can be written as \cite{pais18}:
\begin{eqnarray}
{\cal L} &=& \sum_{j=n,p,d,t,h,\alpha} {\cal L}_{j}                            
+ {\cal L}_{\sigma} + {\cal L}_{\omega} 
+{\cal L}_{\rho} + {\cal L}_{\omega\rho}  + {\cal L}_{e}.
\end{eqnarray} 
The nucleonic gas term is given by
\begin{eqnarray}
{\cal L}_j &=& \bar{\psi}\left[\gamma_\mu i D^\mu - m^*\right]\psi
\end{eqnarray}
with
\begin{eqnarray}
i D^\mu&=&i\partial^\mu-g_v\omega^\mu-\frac{g_\rho}{2}{\boldsymbol\tau}_j \cdot \mathbf{b}^\mu \, ,  \\
m^*&=&m-g_s\phi_0 \, ,
\end{eqnarray}
where $m^*$ is the nucleon effective mass, and $m=m_p=m_n$ is the vacuum nucleon mass, taken as 939 MeV. ${\boldsymbol\tau}_j$ is the isospin operator.
$g_s, g_v$ and $g_{\rho}$ are the couplings of the nucleons to the mesons. The meson fields are given by:
\begin{eqnarray}
{\cal L}_\sigma&=&+\frac{1}{2}\left(\partial_{\mu}\phi\partial^{\mu}\phi
-m_s^2 \phi^2 - \frac{1}{3}\kappa \phi^3 -\frac{1}{12}\lambda\phi^4\right),\nonumber\\
{\cal L}_\omega&=&-\frac{1}{4}\Omega_{\mu\nu}\Omega^{\mu\nu}+\frac{1}{2}
m_v^2 \omega_{\mu}\omega^{\mu}, \nonumber \\ 
{\cal L}_\rho&=&-\frac{1}{4}\mathbf B_{\mu\nu}\cdot\mathbf B^{\mu\nu}+\frac{1}{2}
m_\rho^2 \mathbf b_{\mu}\cdot \mathbf b^{\mu}, \nonumber \\ 
{\cal L}_{\omega\rho}&=& g_{\omega\rho} g_\rho^2 g_v^2 \omega_{\mu}\omega^{\mu}\mathbf b_{\nu}\cdot \mathbf b^{\nu}, 
\end{eqnarray}
where
$\Omega_{\mu\nu}=\partial_{\mu}\omega_{\nu}-\partial_{\nu}\omega_{\mu}, $ and $ \mathbf B_{\mu\nu}=\partial_{\mu}\mathbf b_{\nu}-\partial_{\nu} \mathbf b_{\mu}
- g_\rho (\mathbf b_\mu \times \mathbf b_\nu)$.
The electron contribution is defined as
\begin{equation}
\mathcal{L}_e=\bar \psi_e\left[\gamma_\mu\left(i\partial^{\mu}
\right)-m_e\right]\psi_e \, ,
\label{lage}
\end{equation}
with $m_e$ and $e$ the mass and charge of the electron. 
The Lagrangian density term for the fermionic clusters, e.g. $t$ and $h$, is given by
\begin{eqnarray}
{\cal L}_i &=& \bar{\psi}\left[\gamma_\mu i D_i^\mu - M_i^*\right]\psi,
\end{eqnarray}
with  
\begin{equation}
iD^{\mu }_i = i \partial ^{\mu }-g_{v}^i \omega^{\mu }-\frac{g_\rho}{2}{\boldsymbol\tau}_i \cdot \mathbf{b}^\mu ,
\end{equation}
For the deuteron and $\alpha$ clusters, we have
\begin{eqnarray}
\mathcal{L}_{\alpha }&=&\frac{1}{2} (i D^{\mu}_{\alpha} \phi_{\alpha})^*
(i D_{\mu \alpha} \phi_{\alpha})-\frac{1}{2}\phi_{\alpha}^* \pc{M_{\alpha}^*}^2
\phi_{\alpha},\\
\mathcal{L}_{d}&=&\frac{1}{4} (i D^{\mu}_{d} \phi^{\nu}_{d}-
i D^{\nu}_{d} \phi^{\mu}_{d})^*
(i D_{d\mu} \phi_{d\nu}-i D_{d\nu} \phi_{d\mu})\nonumber\\
&&-\frac{1}{2}\phi^{\mu *}_{d} \pc{M_{d}^*}^2 \phi_{d\mu},
\end{eqnarray}
with
\begin{equation*}
iD^{\mu }_i = i \partial ^{\mu }-g_{v}^i \omega^{\mu }
\end{equation*}
where  $g_v^i$ is the  coupling of cluster $i$ to the vector meson $\omega^\mu$ and it is defined  as $g_{v}^i=A_i g_v$ for all clusters.

We generalize the formalism for all the clusters with $2\leq A\le 12$. In the mean-field approximation and for homogeneous matter, the energy
 density of each particle is given by
\begin{eqnarray}
{\cal E}_i=&\dfrac{2S^i+1}{2\pi^2}\int
             k_i^2E_i (f_{i+}(k)+f_{i-}(k))dk_i
  \nonumber\\
&+g_v^i\omega^0\rho_i+g_\rho b_3^{0}I_3^i\rho_i \, ,
\end{eqnarray} 
where $S^i$, $I_3^i$, and $\rho_i$ are the spin, isospin and density of each cluster, respectively. $E_i$ is the single-particle energy, $E_i=\sqrt{k_i^2+M_i^{*2}}$, and $f_{i\pm}$ are the distribution functions for the particles and
antiparticles, given by:
\begin{eqnarray}
f_{i\pm}&=&\frac{1}{\exp[(E_i\mp\nu_i)/T]+\eta},
\end{eqnarray}
with $\eta=1$ for fermions and $\eta=-1$ for bosons, and $\nu_i=\mu_i-g_v^i \omega^0-g_{\rho} I_3^i b_3^0$. $M_i^*$ is the effective mass of each cluster and it is going to be defined next.

The total binding energy of each cluster is given by
    \begin{eqnarray}
B_i=A_i m^*-M_i^* \,, \label{binding}
\end{eqnarray}
with $M_i^*$  the effective mass given by
\begin{eqnarray}
M_i^*&=&A_i m - g_{s}^i\phi_0 - \left(B_i^0 + \delta B_i\right),
\label{meffi2}
\end{eqnarray}
where $B_i^0$ is the cluster binding energy in the vacuum, and $\delta B_i$ is defined as \cite{pais18}
\begin{eqnarray}
\delta
  B_i=\frac{Z_i}{\rho_0}\left(\epsilon_p^*-m\rho_p^*\right)+\frac{N_i}{\rho_0}\left(\epsilon_n^*-m\rho_n^*\right)
  \, ,
\label{deltaB}
\end{eqnarray}

The binding energy shift, $\delta B_i$, takes in-medium effects into
account, and needs to be determined. It is the energetic counterpart of the classical excluded-volume mechanism. Since $\epsilon_j^*$ and $\rho_j^*$, $j=n,p$ are the energy density and density of the gas in the lowest states, defined as
\begin{eqnarray}
\epsilon_j^*&=&\frac{1}{\pi^2}\int_0^{k_{F_j}(\rm gas)} k^2 E_j (f_{j+}(k)+f_{j-}(k)) dk \\
\rho_j^* &=&\frac{1}{\pi^2}\int_{0}^{k_{F_j}(\rm gas)}  k^2 (f_{j+}(k)+f_{j-}(k)) dk \,,
\end{eqnarray}
we avoid double counting because the energy states occupied by the gas are excluded. The binding energy shift, $\delta B_i$, is reducing the total binding energy of the clusters because it is a negative quantity, since the energy density $\epsilon^*$ is smaller than $m\rho^*$. The fact that $(\epsilon^*- m \rho^*)< 0$ is due to the effect of the $\sigma$ meson that binds matter, see Eq. (\ref{meffi2}), so that the energy per particle is smaller than the vacuum mass for the  densities of interest. The contribution of the binding energy shift, $\delta B_i$, to the total binding energy, $B_i$, is very small, as it can be seen from Fig. 2 of Ref. \cite{pais18}, where this formalism was introduced. 

The other quantity that considers in-medium effects is the scalar cluster-meson coupling, $g_{s}^i=x_{s}^i A_i g_s$, which is determined from experimental constraints. We fix $x_{s}^i$ so that in the low-density limit the Virial EoS is reproduced. We obtained \cite{pais18} $x_{s}^i=0.85\pm 0.05$ as good universal scalar cluster-meson coupling, that not only reproduces reasonably well the Virial EoS but also reproduces well data coming from heavy-ion collisions in the high density limit. We will consider this result for the cluster meson couplings throughout this paper.

To construct the equation of state of warm stellar matter for homogeneous matter with light clusters, we define the total baryonic density as 
\begin{eqnarray}
\rho=\rho_p+\rho_n+\sum_{i=cl}A_i\rho_i
  \, ,
\end{eqnarray}
and we fix the global proton fraction, $Y_p$, as
\begin{eqnarray}
Y_p=y_p+\sum_{i=cl}\frac{Z_i}{A_i}y_i
  \, ,
\end{eqnarray}
with $y_i=A_i(\rho_i/\rho)$. Charge neutrality must be imposed, $\rho_e=Y_p\rho$. The light clusters are in chemical equilibrium, and we define the chemical potential of each cluster as 
\begin{eqnarray}
\mu_i=N_i\mu_n+Z_i\mu_p  \, .
\end{eqnarray}

\subsection{CLD with light clusters}

\begin{figure}
 \begin{tabular}{c}
\includegraphics[width=0.45\textwidth]{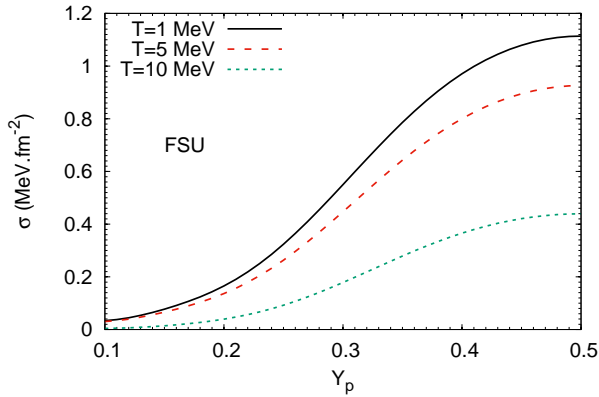} 
 \end{tabular}
  \caption{Surface tension as a function of the global proton fraction for the FSU model and 3 different temperatures.} 
\label{fig0}
\end{figure}

In the compressible liquid drop model (CLD) \cite{pais15}, just like in the coexistence-phase (CP) approximation, matter is divided in two main regions: a high-density phase, constituted by the heavy clusters, and a low-density phase, formed by a background gas of nucleons and light clusters. The equilibrium conditions are obtained by imposing the Gibbs conditions.  The surface tension is both dependent on the  temperature and on the proton fraction, and is obtained within a Thomas-Fermi calculation according to the method discussed in \cite{surface1,surface2}. The specific surface tension parameters for the FSU model, which is used in this work, are given in Ref. \cite{providencia12} and plotted in Fig. \ref{fig0}. The dissolution of the heavy clusters is strongly influenced by the behavior of the surface tension and they are expected to dissolve at smaller densities for smaller proton fractions and larger temperatures.

The main difference with respect to the coexistence-phase approximation method is that the minimization of the free energy is done  only after the surface and Coulomb terms are included. 

The free energy density is given by
\begin{eqnarray}
F&=&fF^I + (1-f)F^{II} + F_e +\varepsilon_{\rm surf} + \varepsilon_{\rm Coul}, \label{totalfree}
\end{eqnarray}
where $F^I$ and $F^{II}$ are the free energy densities of the high- and low-density phases $I$ and $II$, respectively, and $F_e$ is the contribution of the electrons.

Its minimization is done with respect to four variables: the size of the geometric configuration, $r_d$, which gives, just like in the CP case, the condition $\varepsilon_{\rm surf}= 2\varepsilon_{\rm Coul}$ \cite{maruyama05}, the baryonic density in the high-density phase, $\rho^{I}$, the proton density in the high-density phase, $\rho_p^I$, and the volume fraction, $f$, defined as
\begin{eqnarray}
f=\frac{\rho-\rho^{II}}{\rho^{I}-\rho^{II}} \, .
\end{eqnarray}

The equilibrium conditions then become
\begin{eqnarray}
P^I&=&P^{II}-\varepsilon_{\rm surf}\left(\frac{1}{2\alpha}+\frac{1}{2\Phi}\frac{\partial\Phi}{\partial f}-\frac{\rho_p^{II}}{f(1-f)(\rho_p^I-\rho_p^{II})}\right), \nonumber \\
\mu_n^I&=&\mu_n^{II} \, , \nonumber  \\
\mu_p^I&=&\mu_p^{II}-\frac{\varepsilon_{\rm surf}}{f(1-f)(\rho_p^I-\rho_p^{II})} \,,  
 \label{cp}
\end{eqnarray}
with $\alpha=f$ for droplets, rods and slabs, $\alpha=1-f$ for tubes and bubbles. The expression for $\Phi$   depends on the dimension, $D$ and volume fraction, $f$, of the heavy clusters, and is given by \cite{pais15}
\begin{eqnarray}
\Phi=\left\lbrace\begin{array}{cc}
&\left(\frac{2-D \alpha^{1-2/D}}{D-2}+\alpha\right)\frac{1}{D+2} \, ,\quad D=1,3 \\
&\frac{\alpha-1-\ln \alpha}{D+2} \, , \quad D=2 \, .
\end{array} \right. 
\end{eqnarray}
For each phase, the light clusters, which we extend to $A=12$, are in chemical equilibrium, with the
chemical potential of each cluster defined as:
\begin{eqnarray}
\mu_{A_{cl}}^I&=&N\mu_n^I-Z\mu_p^I , \nonumber \\
\mu_{A_{cl}}^{II}&=&N\mu_n^{II}-Z\mu_p^{II} \, , 2\le A_{cl}\le 12 \, ,
\label{mu}
\end{eqnarray}
and charge neutrality must also be imposed:
\begin{eqnarray}
\rho_e&=&Y_p\rho=f\rho_c^I+(1-f)\rho_c^{II} \, ,
\label{rho}
\end{eqnarray}
with $\rho_e$ the electron density and $\rho_c$ the charge density.
Equations (\ref{cp}), (\ref{mu}) and (\ref{rho}) need to be solved self-consistently for the low-energy state to be found.

Let us point out that in the present work we will not consider stellar matter in beta-equilibrium but in conditions appropriate to describe core-collapse supernova matter before and just after the bounce. We, therefore, consider electrically neutral  matter with a fixed proton fraction. The electron contribution to the total Coulomb energy of the droplet configuration leads to the so-called lattice energy, and it is included through the function $\Phi$ defined in Eq. (24) as in Refs. \cite{Ravenhall1983,Maruyama2005}, while deformations of the electron distribution due to the interaction with the nucleus Coulomb field is a higher order effect, which was shown to be negligible for these applications \cite{Watanabe2005}.

\section{Inclusion of ``exotic" clusters}\label{sec:Exotic}

\begin{figure*}
 \begin{tabular}{c}
\includegraphics[width=0.9\textwidth]{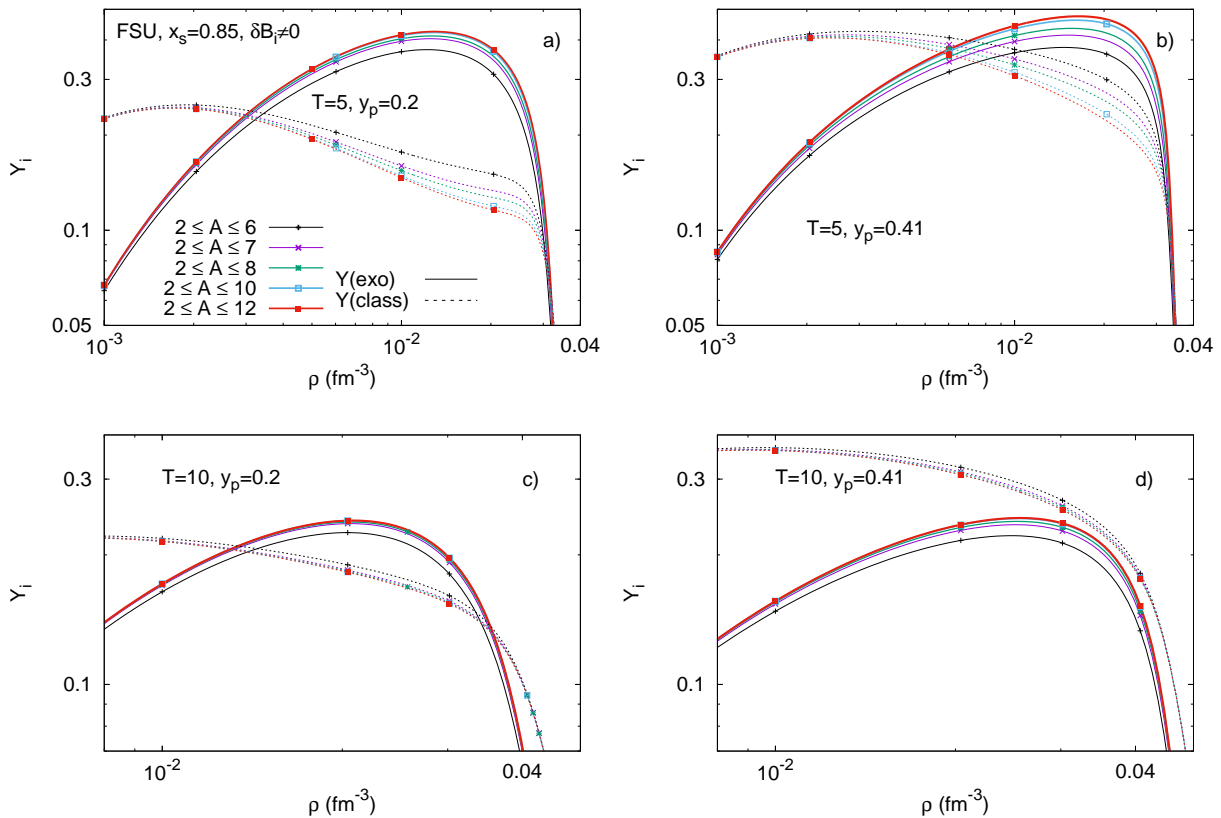} 
 \end{tabular}
  \caption{Mass fraction of the ``classical" light clusters, i.e. $^4$He, $^3$He, $^3$H, $^2$H, (dashed), and mass fraction of the ``exotic" ones (solid), for FSU, $T=5$ (top) and 10 MeV (bottom) , taking $y_p=0.2$ (left), and $y_p=0.41$ (right), for several calculations with different choices for the maximum baryonic number allowed for the light clusters.} 
\label{fig1}
\end{figure*}

In the present work, we include light clusters as point like
particles, in parallel to the neutrons and protons, and heavy clusters
that result from a CLD model approach.  

The most commonly used supernova equations of state \cite{Lattimer,HShen,shen2010} assume that at each thermodynamic condition dense matter is composed of a dominant heavy cluster immersed in a gas of free electrons, protons, neutrons and $\alpha$ particles.
It is, however, known from Nuclear Statistical Equilibrium calculations \cite{Hempel,Raduta} that, in principle, other light nuclear species, different from $\alpha$ particles, can be formed, including loosely bound and unstable nuclei, even if their abundance decreases with increasing baryon number.

In this section, we show the effect of including light clusters of different atomic $Z$ and baryonic $A$ number, in order to determine how far we have to go in $A$ to get convergent results.

We classify the following four clusters, deuteron, triton, helion and
$\alpha$, as ``classical" light clusters, because they have already been considered in the composition of dense matter at finite temperature by different authors \cite{typel10,roepke11,providencia12}.
The bound nuclear species with $4\leq A\leq 12$ will be called  ``exotic" light clusters. 
The formalism is developed within the FSU model \cite{FSU}, a
model that reproduces well the properties of nuclear matter at
saturation and sub-saturation densities, describing, therefore,
reasonably well the inner crust of stars. Although it can not produce
2 solar-mass neutron stars, this problem can be overcome by including an
extra potential above the saturation density that prevents the effective mass
from decreasing, making the EoS harder \cite{Maslov}.

\begin{figure*}
 \begin{tabular}{c}
\includegraphics[width=0.9\textwidth]{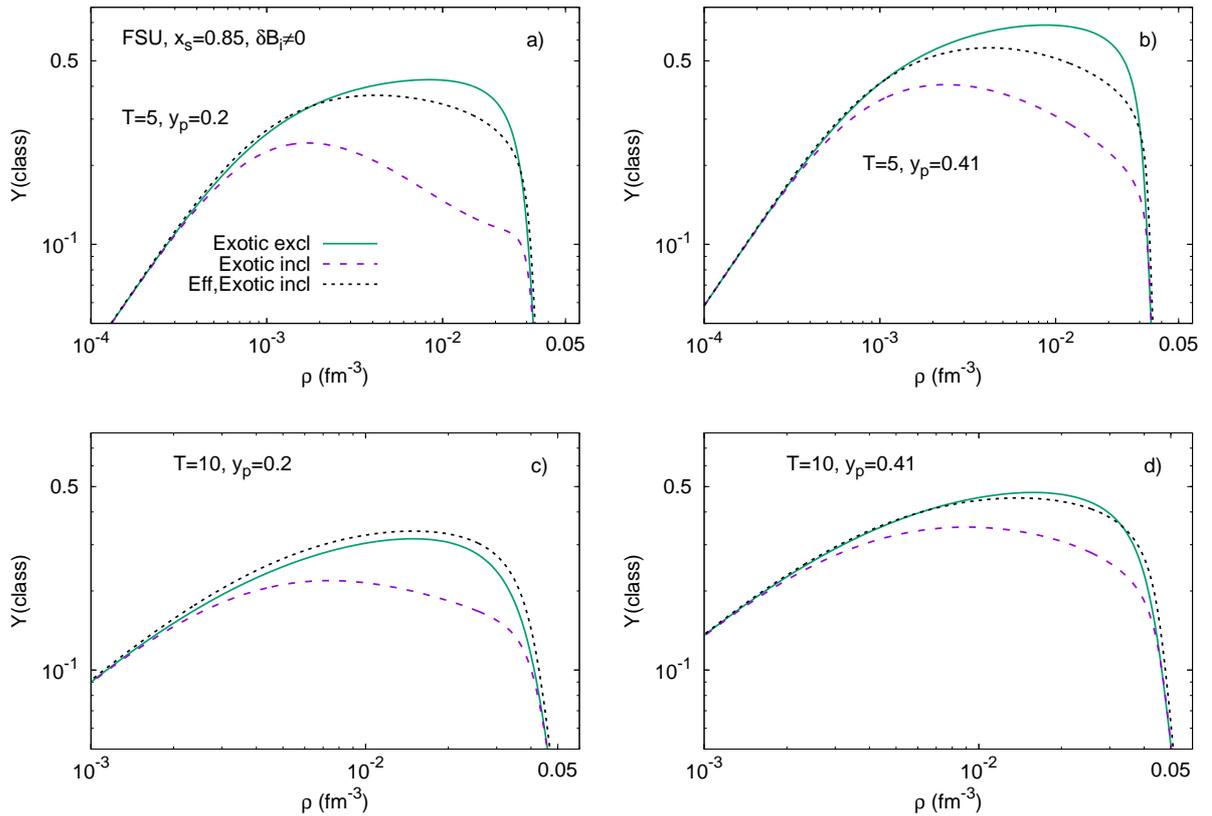} 
 \end{tabular}
  \caption{Mass fraction of the ``classical" light clusters, i.e. $^4$He, $^3$He, $^3$H, $^2$H, when the ``exotic" clusters are included (dashed blue) and when they are excluded (solid green) in the calculation, for FSU and $T=5$ (top) and 10 MeV (bottom), with $y_p=0.2$ (left),  and $y_p=0.41$ (right) with $A\leq 12$. The  dotted black line shows the mass fraction of the ``classical" clusters, taking into account the decays (see text). } 
\label{fig2}
\end{figure*}

Fig.~\ref{fig1} shows the mass fraction of ``exotic" and ``classical" clusters considering
different calculations where all experimentally known nuclear species \cite{ame2016} were included up to a maximum cluster baryonic number which is varied from $A_{max}= 6$ to
$A_{max}=$ 12. The mass fraction of clusters is defined as 
\begin{eqnarray}
Y_i &=&A_i \frac{\rho_i(A_i,Z_i)}{\rho} \nonumber \\ 
Y_A&=&\sum_Z Y_i(Z,A_i=A) \nonumber \\ 
Y_Z&=&\sum_A Y_i(Z_i=Z,A)\nonumber \\
Y_I&=&\sum_A Y_i(I_i=I,A)\nonumber \\
Y_{light}&=&\sum_{i=2}^{A_{max}} Y_i \nonumber \\
Y_{class}&=&Y_d+Y_t+Y_h+Y_\alpha. \nonumber \\
Y_{exo}&=&Y_{light}-Y_{class}
\end{eqnarray}
throughout the text. In the above expressions, $I$ is the isospin projection of each cluster and it is defined as $I=(Z-N)/2$.

The sum in the expression of $Y_{exo}$ is limited to the ``exotic" clusters, that is, it excludes $^2$H,$^3$H,$^3$He, and $^4$He. Note that our approximation of considering clusters as point-like particles 
would not be adequate for heavy clusters, which we treat separately by including their spatial density distribution.  Since the role of light clusters is most important in the presence of heavy clusters, their fraction is always small and we believe that their spatial extension
does not play an important role. A similar approach was undertaken in \cite{typel10}. The influence of the medium in the light clusters is taken into account through the shift on the binding energy and the couplings of the clusters to the mesons.

As we can see from Fig.~\ref{fig1}, results taking $A\le$ 10 and $A\le$ 12 do not change much  the
  total cluster distribution $Y_{light}$, and, therefore, in the following we will not consider clusters with
$A>12$. The largest contribution of the ``exotic" clusters occurs for intermediate densities, when the total distribution of clusters has a peak, as we will see next.

\section{Light clusters with $A\leq12$}\label{sec:A12}

In the following, we discuss the role of the ``exotic" clusters and
the effect of including them in the calculation of warm
non-homogeneous matter. In particular, we will discuss a) their relative
abundance with respect to the ``classical" clusters, b) which clusters
give a larger contribution, and c) we will define ``effective classical" cluster fractions, 
that can be compared to experimental cluster yields measured in heavy ion collisions.

\subsection{How important are the ``exotic" clusters?}
Taking into account the results of the previous section, in the
following, we include in the calculations, besides the ``classical"
light clusters, the ``exotic" light clusters with $A\le 12$, and study
their contribution in detail.

Fig.~\ref{fig2} shows the fractions of ``classical"  light clusters for
$T=5$ and 10 MeV, and proton fractions 0.2 and 0.41, considering two
calculations: including or excluding ``exotic" clusters  with $A\le 12$. The solid black line shows the total mass fraction of the effective ``classical" light clusters, which takes into account the decay modes of the ``exotic" clusters into the ``classical" ones, and it will be discussed in more detail in Section \ref{Sec:decay}.
The inclusion of the ``exotic" clusters has no effect on the low
density distribution of the ``classical" clusters close to the cluster
onset, neither on the cluster distribution close to the melting densities.  The largest differences
occur at the maximum of the cluster distribution, and indicate
that for these densities, a larger
number of degrees of freedom contribute. The main implications of the
cluster mass distribution are related with the
contribution that clusters may give to transport properties of matter.

\subsection{Which clusters are the most abundant?}

\begin{figure}
 \begin{tabular}{c}
\includegraphics[width=0.45\textwidth]{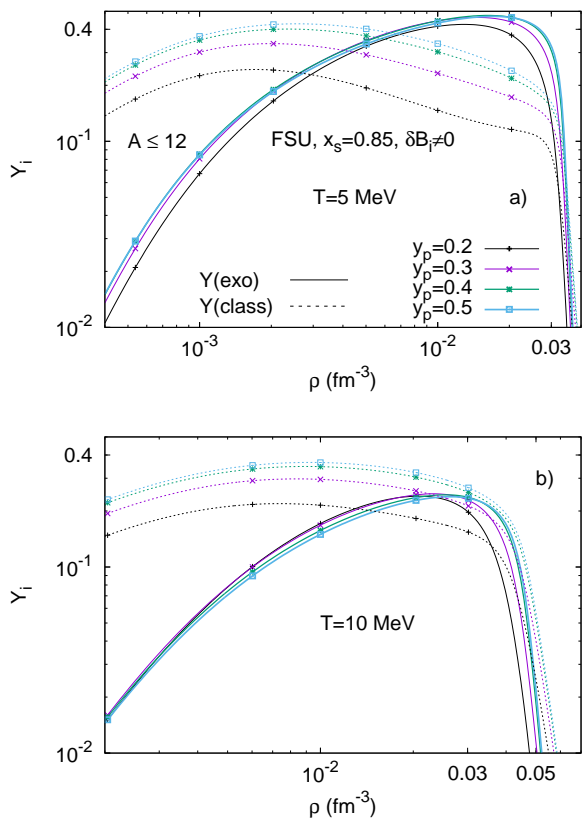} 
 \end{tabular}
  \caption{Mass fraction of the ``classical" light clusters, i.e. $^4$He
    ,$^3$He, $^3$H, $^2$H, (dashed), and mass fraction of the ``exotic" clusters (solid), for the FSU model, $y_p=0.2-0.5$ with $T=5$ (top) and 10 MeV (bottom).} 
\label{fig3}
\end{figure}

\begin{figure*}
 \begin{tabular}{c}\includegraphics[width=0.95\textwidth]{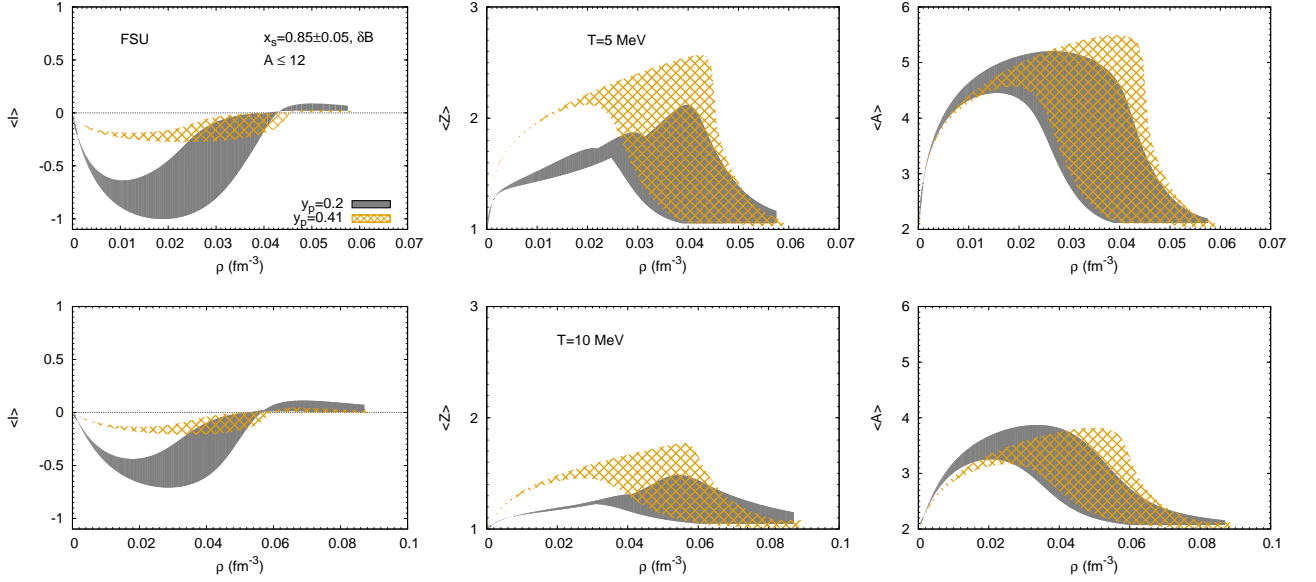} 
 \end{tabular}
  \caption{Average isospin (left), average charge content (middle), and average number of nucleons (right) of the light clusters for $T=5$ MeV (top) and $T=10$ MeV (bottom), with $y_p=0.2$ (solid grey region) and 0.41 (crossed orange region) in a calculation for FSU with $x_s=0.85\pm 0.05$.} 
\label{fig4}
\end{figure*}

We next study the effect of  temperature and proton-neutron matter asymmetry on the
abundances of the light clusters. 
In Fig.~\ref{fig3}, the mass fractions of the ``classical" and ``exotic" light clusters  are
plotted for several proton fractions and $T=5$ and 10 MeV.

The ``exotic" clusters do not play a role at small densities, close and
above the onset density of the ``classical" light clusters. A similar
conclusion is drawn at the transition to homogeneous matter: the
``classical" light clusters determine this transition. However, at the
maximum of the clusters fraction distributions, the ``exotic" clusters
are more abundant if the temperature is not too high. At the  maximum
distribution, the difference between ``exotic" and ``classical"
clusters increases as the proton fraction decreases. The relative contribution of the
``exotic" clusters becomes more important for very asymmetric matter and
low temperatures. These differences have already disappeared for
$T\sim 10$ MeV.

Table~\ref{tab1} shows the five most abundant clusters at five fixed densities, for $T=5$ and 10 MeV and proton fractions of 0.2 and 0.41. At low densities, $^2$H is always the most abundant cluster. This tendency increases to larger densities and for larger temperatures, together with more symmetric matter: for $T=10$ MeV and $y_p=0.41$, $^2$H is the most abundant for all densities. The largest and most asymmetric cluster within the five most abundant clusters is $^7$He, and occurs for the lowest temperature considered, the smallest $y_p$, and at the two largest densities included in the table, $10^{-2}$ and $2\times 10^{-2}$ fm$^{-3}$, though its abundance does not reach 5$\%$. However, it is clear from the table that, at $T=5$ MeV and $y_p=0.41$, the two most abundant clusters concentrate $30-35\%$ of the distribution, with a fast reduction to $10-15\%$ to the next three most abundants ones. For $y_p=0.2$, the distribution is more uniform: the two most abundants correspond to $15-20\%$ of the total distribution, while the less abundant ones to $10-16\%$. Temperature and a smaller proton fraction turn the distribution more uniform.

Other important conclusions can be inferred from the table: i) For $T=10$ MeV and $y_p=0.41$, the heaviest nucleus is $^5$He, with isospin $-1/2$. The heaviest one with largest isospin magnitude is $^4$H ($I=-1$). Reducing the proton fraction to 0.2, $^6$H is the most massive and most asymmetric ($I=-2$), but occurs only with a $3\%$ abundancy. $^4$H and $^5$He are the ``exotic" clusters with the most important contribution. ii) For $T=5$ MeV and $y_p=0.41$, there is no cluster with $A>5$. However, for $y_p=0.2$, $^7$He ($I=-3/2$) has a non-negligible contribution, close to $5\%$. We conclude that the ``exotic" clusters have a more important role at lower temperatures and larger proton-neutron asymmetries. Moreover, the results of Table~\ref{tab1} seem to indicate that it is enough to consider a small subset of ``exotic" clusters with $A\leq 7$ and $|I|\leq\{3/2,2\}$.

\begin{table*}
\caption{The five most abundant light clusters, taken at five fixed total
  densities, $\rho$, for $T=5, 10$ MeV and $y_p=0.2$ and  0.41.}
   \begin{tabular}{lccccc}
\hline
\hline
\multicolumn{6}{c}{$y_p=0.2, T=5$ MeV}\\
\hline
$\rho$ $[$fm$^{-3}]$ &\multicolumn{5}{c}{$^AX$, $Y(^AX)[\%]$}\\
$5\times 10^{-4}$ & $^2$H, $9.50\%$ & $^3$H, $3.86\%$ & $^4$He, $2.18\%$ & $^5$He, $0.82\%$ & $^4$H, $0.71\%$\\
$1\times 10^{-3}$ & $^2$H, $9.94\%$ & $^3$H, $7.08\%$ & $^4$He, $4.81\%$ & $^5$He, $3.15\%$ & $^4$H, $2.29\%$\\
$5\times 10^{-3}$ & $^5$He, $9.69\%$ & $^3$H, $9.24\%$ & $^4$H, $8.40\%$ & $^4$He, $5.27\%$ & $^2$H, $4.62\%$\\
$1\times 10^{-2}$ & $^5$He, $9.90\%$ & $^4$H, $8.48\%$ & $^3$H, $7.34\%$ & $^7$He, $4.86\%$ & $^4$He, $4.25\%$\\
$2\times 10^{-2}$ & $^5$He, $9.85\%$ & $^4$H, $5.44\%$ & $^3$H, $4.94\%$ & $^4$He, $4.46\%$ & $^7$He, $4.38\%$\\
\hline
\multicolumn{6}{c}{$y_p=0.41, T=5$ MeV}\\
\hline
$\rho$ $[$fm$^{-3}]$ &\multicolumn{5}{c}{$^AX$, $Y(^AX)[\%]$}\\
$5\times 10^{-4}$ & $^2$H, $14.21\%$ & $^4$He, $4.91\%$ & $^3$H, $3.67\%$ & $^3$He, $1.94\%$ & $^5$He, $1.16\%$\\
$1\times 10^{-3}$ & $^2$H, $15.19\%$ & $^4$He, $11.23\%$ & $^3$H, $6.06\%$ & $^5$He, $4.10\%$ & $^3$He, $2.87\%$\\
$5\times 10^{-3}$ & $^4$He, $19.19\%$ & $^5$He, $14.17\%$ & $^2$H, $8.80\%$ & $^3$H, $7.09\%$ & $^5$Li, $4.51\%$\\
$1\times 10^{-2}$ & $^4$He, $17.42\%$ & $^5$He, $15.41\%$ & $^2$H, $5.87\%$ & $^3$H, $5.68\%$ & $^5$Li, $4.63\%$\\
$2\times 10^{-2}$ & $^4$He, $13.78\%$ & $^5$He, $12.98\%$ & $^5$Li, $4.36\%$ & $^3$H, $3.73\%$ & $^2$H, $3.70\%$\\
\hline
\multicolumn{6}{c}{$y_p=0.2, T=10$ MeV}\\
\hline
$\rho$ $[$fm$^{-3}]$ &\multicolumn{5}{c}{$^AX$, $Y(^AX)[\%]$}\\
$1\times 10^{-3}$ & $^2$H, $7.42\%$ & $^3$H, $1.20\%$ & $^3$He, $0.23\%$ & $^4$H, $0.22\%$ & $^4$He, $0.09\%$\\
$5\times 10^{-3}$ & $^2$H, $12.02\%$ & $^3$H, $7.16\%$ & $^4$H, $4.82\%$ & $^5$He, $1.29\%$ & $^4$He, $1.15\%$\\
$1\times 10^{-2}$ & $^2$H, $10.10\%$ & $^4$H, $8.92\%$ & $^3$H, $8.91\%$ & $^5$He, $2.76\%$ & $^5$H, $1.80\%$\\
$2\times 10^{-2}$ & $^4$H, $10.32\%$ & $^3$H, $8.22\%$ & $^2$H, $7.48\%$ & $^5$He, $3.95\%$ & $^6$H, $3.03\%$\\
$3\times 10^{-2}$ & $^4$H, $7.70\%$ & $^3$H, $6.46\%$ & $^2$H, $6.17\%$ & $^5$He, $3.98\%$ & $^4$He, $2.01\%$\\
\hline
\multicolumn{6}{c}{$y_p=0.41, T=10$ MeV}\\
\hline
$\rho$ $[$fm$^{-3}]$ &\multicolumn{5}{c}{$^AX$, $Y(^AX)[\%]$}\\
$1\times 10^{-3}$ & $^2$H, $11.26\%$ & $^3$H, $1.26\%$ & $^3$He, $0.76\%$ & $^4$He, $0.20\%$ & $^4$H, $0.16\%$\\
$5\times 10^{-3}$ & $^2$H, $19.31\%$ & $^3$H, $6.76\%$ & $^3$He, $3.62\%$ & $^4$He, $2.97\%$ & $^4$H, $2.68\%$\\
$1\times 10^{-2}$ & $^2$H, $17.36\%$ & $^3$H, $8.42\%$ & $^4$He, $4.89\%$ & $^4$H, $4.63\%$ & $^5$He, $4.47\%$\\
$2\times 10^{-2}$ & $^2$H, $13.12\%$ & $^3$H, $7.93\%$ & $^5$He, $6.64\%$ & $^4$He, $5.80\%$ & $^4$H, $5.48\%$\\
$3\times 10^{-2}$ & $^2$H, $10.12\%$ & $^5$He, $6.55\%$ & $^3$H, $6.48\%$ & $^4$He, $5.39\%$ & $^4$H, $4.77\%$\\
\hline
\hline
\end{tabular}
\label{tab1}
\end{table*}

In order to   establish the role of isospin, charge and mass, 
we plot in Fig.~\ref{fig4}, the average isospin of the light
clusters, $<I>$, the average charge $<Z>$, and the average number of
nucleons in the light clusters $<A>$, given by 
\begin{eqnarray}
<I>&=&\frac{\sum_i I_i \rho_i}{\sum_i \rho_i} \\
<Z>&=&\frac{\sum_i Z_i \rho_i}{\sum_i \rho_i} \\
<A>&=&\frac{\sum_i A_i \rho_i}{\sum_i \rho_i} 
\end{eqnarray}
as a function of density. The  regions shown in  Fig.~\ref{fig4}
were obtained allowing for the coupling of the light clusters to the
scalar field to vary in the range $0.8<x_s<0.9$.

 We first discuss the role of isospin. In the left panels of
  Fig.~\ref{fig4}, we plot for two temperatures, $T=5$ and 10 MeV,
  and two proton fractions, $y_p=0.2$ and 0.41, the average isospin cluster.  As expected, for neutron
  rich matter, the clusters that most contribute are neutron rich, and
  the more neutron rich matter is, the larger the fraction of
  clusters with a negative isospin. For $y_p=0.2$ and $T=5$ MeV,  the presence of
clusters that have at least $N-Z=2$ is large, although temperature
  reduces strongly this effect. We, therefore, conclude that 
it  is important to include in the calculation of very asymmetric matter
  exotic neutron rich clusters. Even for $T=10$ MeV, the maximum of
  the average  isospin is above $-\frac{1}{2}$ for $y_p=0.2$.

In the middle panels of  Fig.~\ref{fig4}, the average charge of
the light clusters has been plotted. We consider the same two proton fractions
and temperatures, as before.  It is seen that the presence of clusters with a large charge, i.e $Z>2$,
is more important in symmetric matter and low temperature. In particular, for the proton
fraction $y_p=0.41$,  the effect of clusters with $Z>2$ at $T=5$ MeV  is
non-negligible, while for  $y_p=0.2$ or $T=10$ MeV, their role is small. The
larger the value of $x_s$, the larger the contribution of clusters with
$Z>2$.

 We finally refer to the role of the light cluster mass. The
  average mass number of the light clusters is shown in the right panels of  Fig.~\ref{fig4}. This quantity is essentially not  affected by the proton fraction, but it is sentitive to the temperature:
  the larger the $T$, the smaller the contribution from the most massive
clusters. For $T=10$ MeV, the maximum mass average of clusters is
$\sim 4$, while for $T=5$ MeV, this value raises to $\sim
5$. The fraction $x_s$ also has a noticeable effect: larger
values favor more massive clusters, because it introduces a larger attraction.


\subsection{Consideration of decay modes}\label{Sec:decay}

Having as an objective the comparison of the cluster abundances
within our model with the experimental  data, we will calculate
equilibrium constants as defined in  \cite{qin12}. When considering ``exotic"
clusters we should keep in mind that many of them are unstable in
vacuum, and therefore, in heavy ion collisions, they will decay before reaching the
detectors. We, therefore, introduce effective cluster mass fractions and
densities that take this effect into account. Specifically, we sum up
all the cluster mass fractions that decay into a given stable light cluster, thus mimicking the final yield 
that is measured in a heavy ion experiment after secondary decay. Of the four ``classical" light clusters, only $^3$H and $^4$He have effective densities and mass fractions because none of the ``exotic" clusters decay into $^2$H no $^3$He.

The following decay modes are going to be considered \cite{ame2016} (for simplicity, the leptons emitted in the decay are not specified):
\begin{eqnarray*}
^5{\rm He} &\longrightarrow & ^4{\rm He} + {\rm n} \\
^4{\rm H} &\longrightarrow & ^3{\rm H} + {\rm n} \\
^7{\rm He} &\longrightarrow & ^6{\rm Li} +{\rm n} \\
^6{\rm H} &\longrightarrow & ^3{\rm H} + 3{\rm n} \\
^5{\rm H} &\longrightarrow & ^3{\rm H} + 2{\rm n} \\
^5{\rm Li} &\longrightarrow & ^4{\rm He} + {\rm p} \\
^8{\rm Be} &\longrightarrow & 2(^4{\rm He}) \\
^7{\rm Be} &\longrightarrow & ^7{\rm Li} \\
^9{\rm He} &\longrightarrow & 2(^4{\rm He}) + {\rm n} \\
^7{\rm H} &\longrightarrow & ^3{\rm H} +4{\rm n} 
\end{eqnarray*}
\begin{figure*}
 \begin{tabular}{c}
\includegraphics[width=0.9\textwidth]{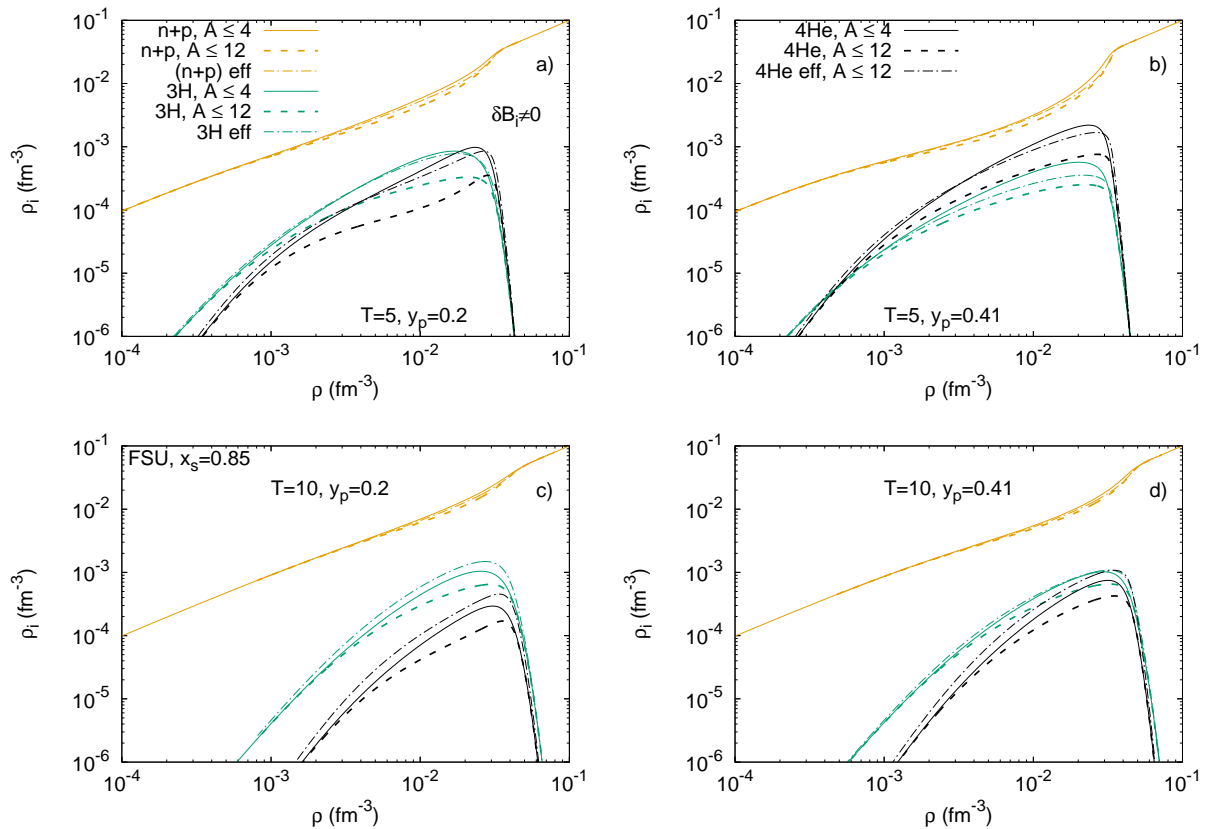} 
 \end{tabular}
  \caption{Effective densities (dash-dotted lines) of free nucleons (orange), tritons (green), and $\alpha$ (black), compared to their primary (without the contribution of secondary decay, see text) densities (dashed lines) in a calculation with $A\leq 12$,  as a function of the total density, considering $T=5$ (top) and 10 MeV (bottom), and taking $y_p=0.2$ (left) and 0.41 (right). A calculation with $A\leq 4$ is also shown (solid lines).} 
\label{fig5}
\end{figure*}
because these are the most abundant clusters, i.e., have a mass fraction of $Y_i>10^{-2}$.
Considering these decays, we define the following effective densities, $\widetilde{\rho_i}$, as:
\begin{eqnarray}
\widetilde{\rho}_{^4{\rm He}}&=&\rho_{^4{\rm He}}+\rho_{^5{\rm He}}
                           +\rho_{^5{\rm Li}} +2\rho_{^8{\rm Be}} +2\rho_{^9{\rm He}}  \nonumber \\
\widetilde{\rho}_{^3{\rm H}}&=&\rho_{^3{\rm H}}+\rho_{^4{\rm H}}+\rho_{^5{\rm H}}+\rho_{^6{\rm H}}+\rho_{^7{\rm H}} \nonumber  \\
\widetilde{\rho}_{^6{\rm Li}}&=&\rho_{^6{\rm Li}}+\rho_{^7{\rm He}} \nonumber  \\ 
\widetilde{\rho}_{^7{\rm Li}}&=&\rho_{^7{\rm Li}}+\rho_{^7{\rm Be}} \nonumber \\
\widetilde{\rho}_{{\rm n}}&=&\rho_{{\rm n}}+\rho_{^5{\rm He}}+\rho_{^4{\rm H}}+\rho_{^7{\rm He}}+3\rho_{^6{\rm H}}\nonumber \\ &&+2\rho_{^5{\rm H}}+\rho_{^9{\rm He}}+4\rho_{^7{\rm H}} \nonumber \\
\widetilde{\rho}_{{\rm p}}&=&\rho_{\rm p}+\rho_{^5{\rm Li}} 
\end{eqnarray}

In Fig.~\ref{fig5},  the $\alpha$, triton and free nucleons (neutrons and protons) densities are shown for two different calculations: a) the effective (dash-dotted lines),  and primary (dashed lines) cluster  densities,  taking $A\leq 12$; and b)  the primary cluster densities (solid lines), taking $A\leq 4$. 
 From the first calculation, we immediately  conclude that the
``exotic" clusters play a non-negligible role at intermediate densities.
 Comparing both calculations, i.e.  the distribution of light
  clusters, with or without the exotics, it is clear that there are
  differences: at the peak of the distribution, the mass fractions without the ``exotic" clusters may be  more abundant, if the temperature is not too high and the proton
  fraction is not too small, but at smaller and larger densities, the
  effective  ``classical" light cluster are more abundant. We can then say that including the ``classical" light clusters only takes
  into account, in a reasonable way, the distribution of light clusters
  with $A\le 12$. However, we will next verify that, in fact, the equilibrium
  constants are affected.

\begin{figure}
 \begin{tabular}{c}
\includegraphics[width=0.45\textwidth]{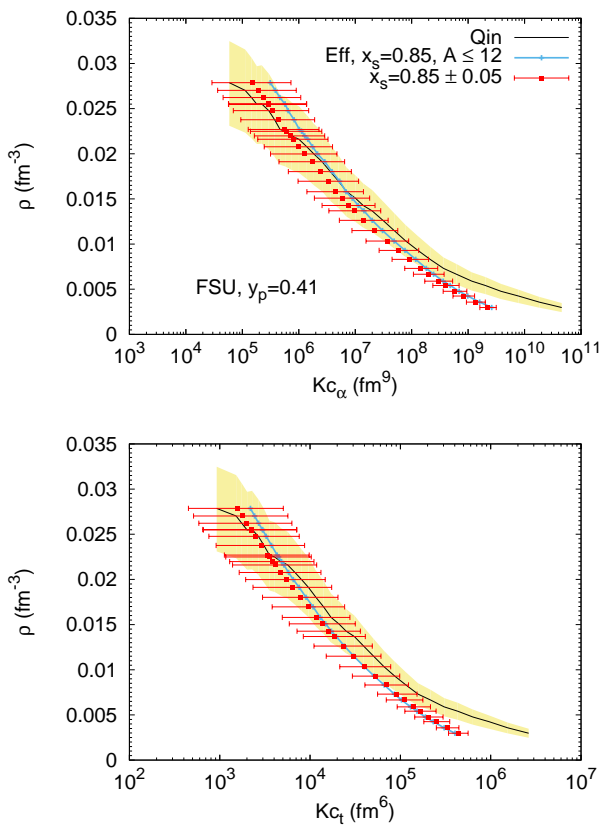} 
 \end{tabular}
  \caption{Chemical equilibrium constants of $\alpha$ (top), and triton (bottom) for FSU, and $y_p=0.41$, and for the universal $g_{s_j}=(0.85\pm 0.05)A_j g_s$ fitting, from a calculation with only the four ``classical" light clusters (red with arrow bars), and a calculation with $A\leq 12$, taking the effective densities, and $x_s=0.85$ (cian/grey thick line). The experimental results of Qin et al. \cite{qin12} (yellow/grey shaded region) are also shown.}
 \label{fig6}
\end{figure}

In Fig.~\ref{fig6}, we have calculated the equilibrium constants, $K_c$, for the $\alpha$ and
triton clusters considering
the two following cases:  a) the  calculation contains  all clusters with
$A\leq 12$, and the $K_c$ are determined for  the corresponding effective distributions (magenta), which we call effective equilibrium constants; b) the  calculation contains  only the four ``classical" clusters, and the $K_c$ are calculated as in Ref.~\cite{pais18} (red).
These distributions are compared  with the experimental
data of Qin et al. \cite{qin12}.
When we compare these two calculations, we see that, for the same
temperature and density, the effective equilibrium constants become larger, as it
might be expected, since besides the true distributions, there is a
large number of other  channels that contribute.

\section{Compressible liquid drop (CLD) calculation with $A\leq 12$}\label{sec:CLD}

\begin{figure}
 \begin{tabular}{c}
\includegraphics[width=0.45\textwidth]{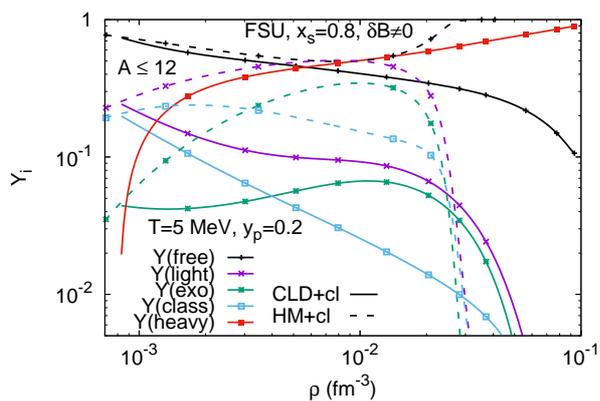} 
 \end{tabular}
  \caption{Total fraction of  free particles (black),  light clusters (magenta), ``exotic" light clusters (green), and classical light clusters (cyan), for a CLD (solid) and HM (dashed) calculations. In both calculations we are including $\delta B$ and $A \leq 12$. The heavy cluster (red) from a CLD calculation is also shown. The results are for FSU, $y_p=0.2$, $x_s=0.8$, and $T=5$ MeV.} 
\label{fig7}
\end{figure}

\begin{figure}
 \begin{tabular}{c}
\includegraphics[width=0.45\textwidth]{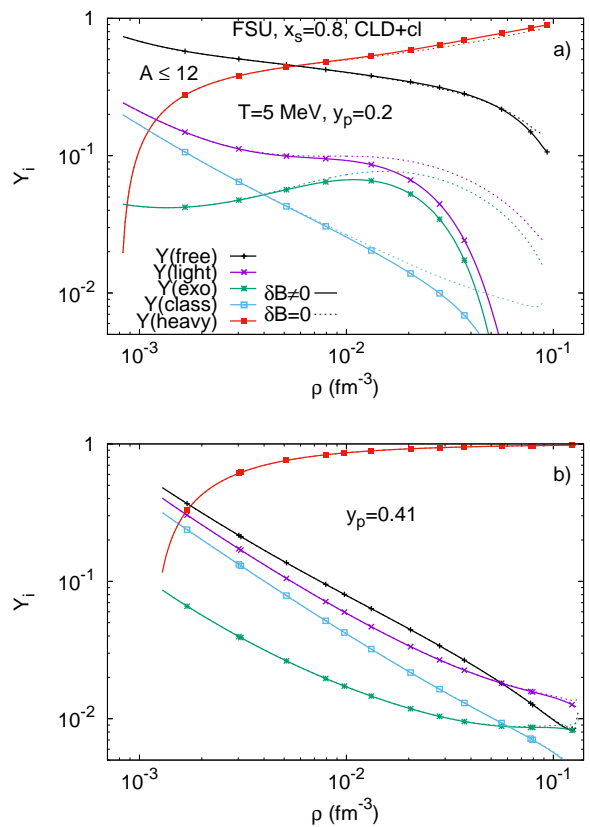} 
 \end{tabular}
  \caption{Total fraction of  free particles (black),  light clusters (magenta), ``exotic" light clusters (green), and classical light clusters (cyan), for a CLD with (solid) and without $\delta B$ (dotted). The heavy cluster (red) is also shown. The results are for FSU, $T=5$ MeV, and $x_s=0.8$, for $y_p=0.2$ (top) and $y_p=0.41$ (bottom). In both calculations we are taking $A \leq 12$.} 
\label{fig8}
\end{figure}

\begin{figure*}
 \begin{tabular}{c}
\includegraphics[width=0.95\textwidth]{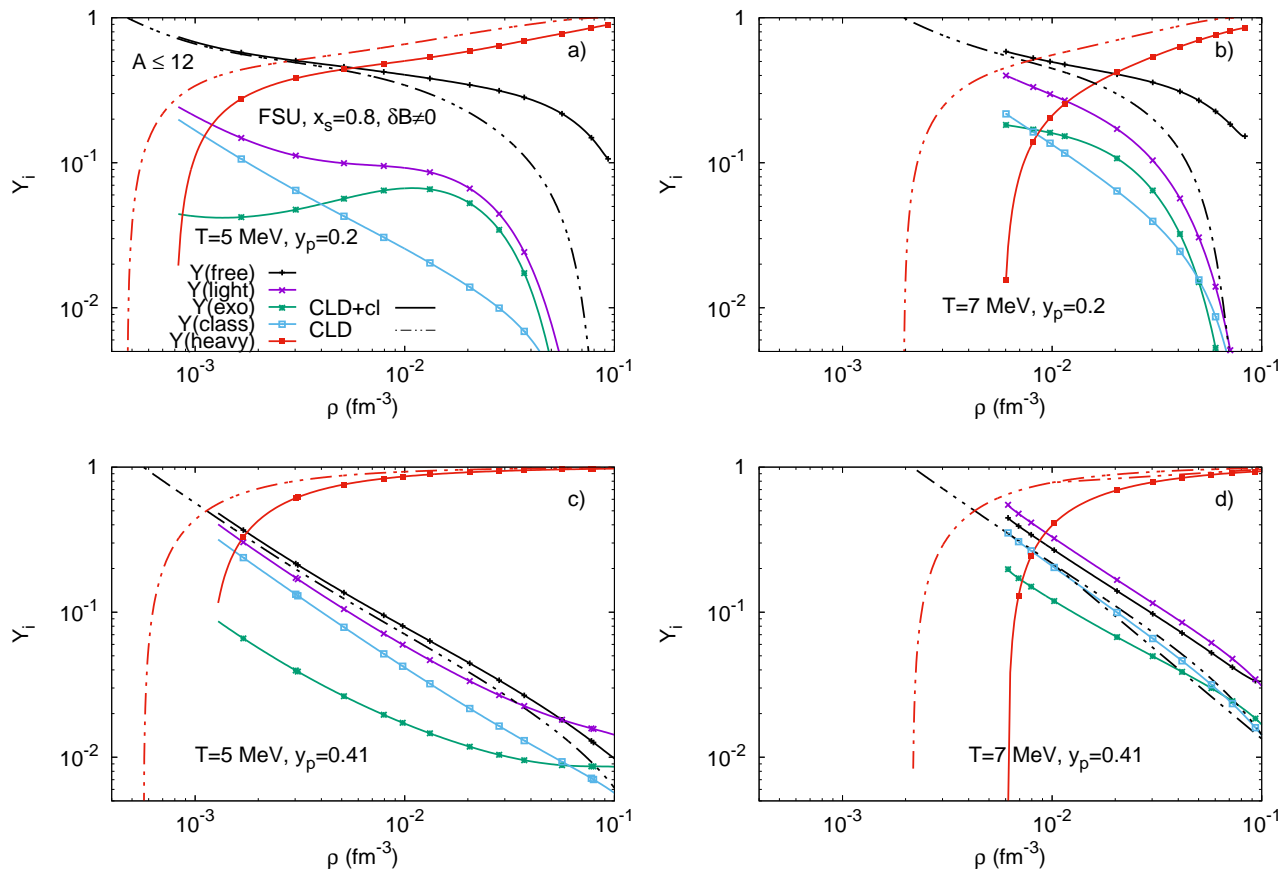} 
 \end{tabular}
  \caption{Total fraction of  free particles (black),  light clusters (magenta), ``exotic" light clusters (green), classical light clusters (cyan), and heavy cluster (red) for a CLD with (solid) and without light clusters (cl) (dash-dotted) calculations. The results are for FSU, $x_s=0.8$, with $y_p=0.2$ (top), and 0.41 (bottom), for $T=5$ (left) and $T=7$ (right). In both calculations, we are taking $A \leq 12$.} 
\label{fig9}
\end{figure*}

\begin{figure*}
 \begin{tabular}{c}
\includegraphics[width=0.95\textwidth]{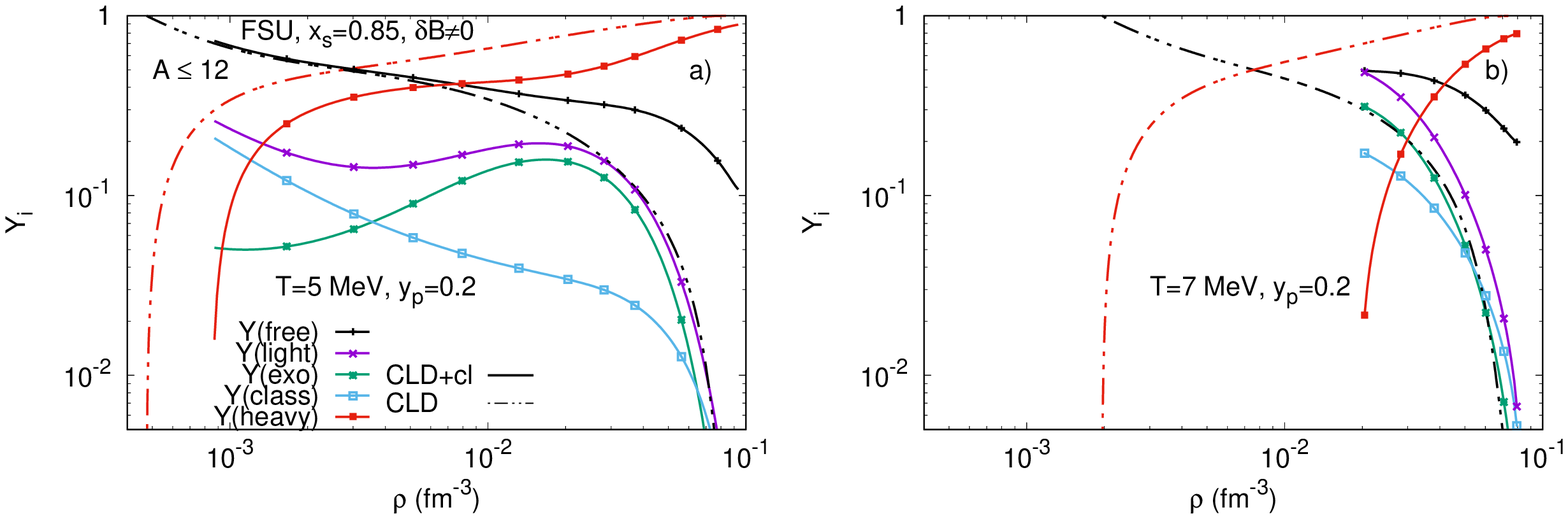}\\
\includegraphics[width=0.95\textwidth]{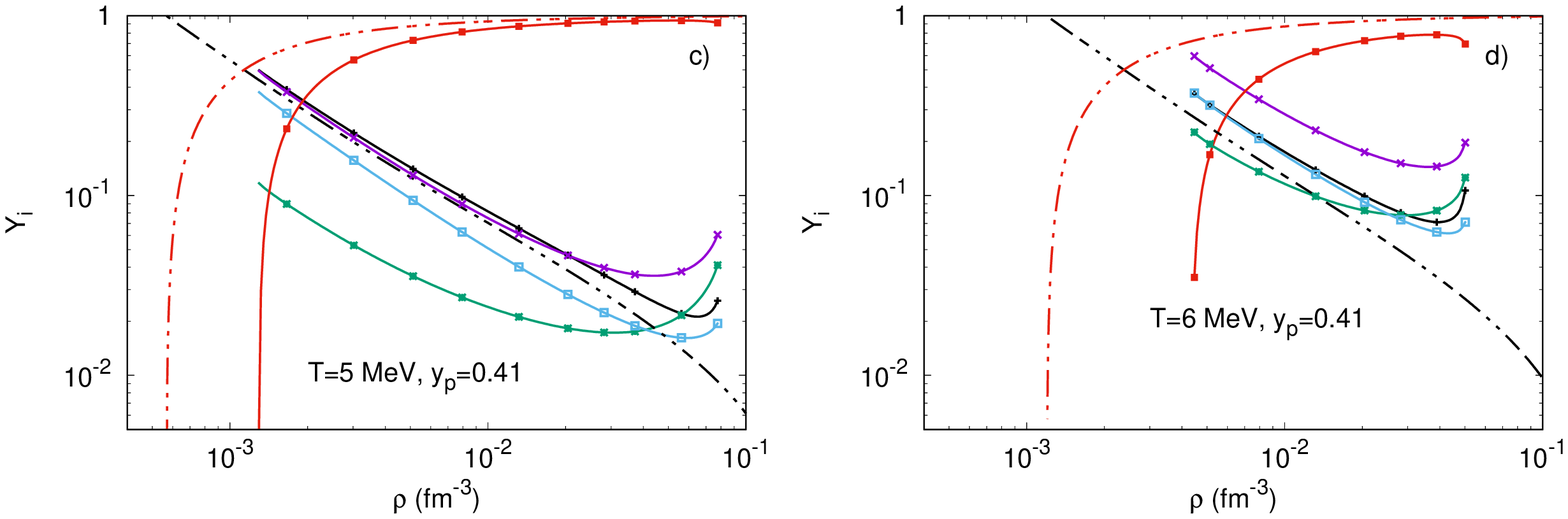} 
 \end{tabular}
  \caption{Total fraction of  free particles (black),  light clusters (magenta), ``exotic" light clusters (green), classical light clusters (cyan), and heavy cluster (red) for a CLD with (solid) and without light clusters (cl) (dash-dotted) calculations. The results are for FSU, and $x_s=0.85$. The top panels are for $y_p=0.2$, with $T=5$ (left) and 7 MeV (right). The bottom panels are for $y_p=0.41$, with $T=5$ (left) and 6 MeV (right). In both calculations, we are taking $A \leq 12$.} 
\label{fig10}
\end{figure*}

Until now, we have considered homogeneous matter (HM) with  light clusters, both the ``classical" and the ``exotic" ones. We next test how
the fraction of heavy clusters (pasta) is affected with the inclusion
of the ``exotic" clusters. For that, we consider a CLD calculation with
light clusters, taking $A\leq 12$, and where the inclusion of the
extra term in the binding energy of the light clusters, $\delta B$, defined  in eq. (\ref{deltaB}), is also considered. In the following, the heavy cluster will always be
calculated in the droplet configuration.
 
In this calculation, we consider the following definitions for the total proton mass fraction, $y_p^{\rm Tot}$, the total neutron mass fraction, $y_n^{\rm Tot}$, and the total mass fraction of a light cluster with $A$ nucleons and $N$ neutrons, $Y_{\rm cl(A,N)}^{\rm Tot}$:
\begin{flushleft}
\begin{eqnarray}
y_p^{\rm Tot}&=&(Y_{p_1}f\rho_1+Y_{p_2}(1-f)\rho_2)/\rho \, , \nonumber \\
y_n^{\rm Tot}&=&(Y_{n_1}f\rho_1+Y_{n_2}(1-f)\rho_2)/\rho \, , \\
Y_{\rm cl(A,N)}^{\rm Tot}&=&(Y_{{\rm cl(A,N)}1}f\rho_1+Y_{{\rm cl(A,N)}2}(1-f)\rho_2)/\rho  \, . \nonumber
\end{eqnarray}
\end{flushleft}
$Y_{i1}$  ($Y_{i2}$) is the particle fraction in the dense phase
1 (gas phase 2), $\rho$ the average baryonic density, and $f$ the volume fraction occupied by the heavy cluster.

It is interesting to observe that, though in the above definition, we allow the presence of light clusters in the whole Wigner-Seitz cell, independent of the density, it turns out that $Y_{{\rm cl(A,N)}1}=0$ for all $(A,N)$, showing that our universal coupling prescription naturally produces the expected excluded volume effect of the dense cluster, here identified with the dense phase 1.
We also  define the total fraction  of free nucleons $Y_{\rm free}$,
the total fraction  of light, ``classical" and ``exotic" clusters,
respectively,  $Y_{\rm light}$, $Y_{\rm class}$, $Y_{\rm exo}$, and the
fraction of nucleons in the heavy cluster $Y_{heavy}$ as
\begin{eqnarray}
Y_{\rm free}&=&(Y_{p_2}+Y_{n_2})(1-f)\rho_2/\rho \, , \\
Y_{\rm light}&=&\sum_{A=2}^{12}Y_{{\rm cl(A,N)}2}(1-f)\rho_2/\rho \, , \nonumber \\
Y_{\rm class}&=&\sum_{A=2}^{4}Y_{{\rm cl(A,N)}2}(1-f)\rho_2/\rho \, , \nonumber \\
Y_{\rm exo}&=&Y_{\rm light}-Y_{\rm class} \, , \\
Y_{\rm heavy}&=&\left(Y_{p_1}+Y_{n_1}+\sum_{A=2}^{12}Y_{{\rm cl(A,N)}1}\right)f \rho_1/\rho \, .
\end{eqnarray}
Finally, let us also define the number of nucleons, $A_{\rm heavy}$, and protons, $Z_{\rm heavy}$, inside
the heavy cluster, and the average density of the heavy cluster, $\rho_{\rm heavy}$,
\begin{eqnarray}
A_{\rm heavy}&=&\frac{4\pi R_d^3}{3}(\rho_1-\rho_2(Y_{p_2}+Y_{n_2})) \, , \\
Z_{\rm heavy}&=&\frac{4\pi R_d^3}{3}(\rho_1Y_{p_1}-\rho_2 Y_{p_2}) \, , \\
\rho_{\rm heavy}&=&Y_{\rm heavy}\rho /A_{\rm heavy} \, . \label{rhoCLD}
\end{eqnarray}

\begin{figure}
 \begin{tabular}{c}
\includegraphics[width=0.45\textwidth]{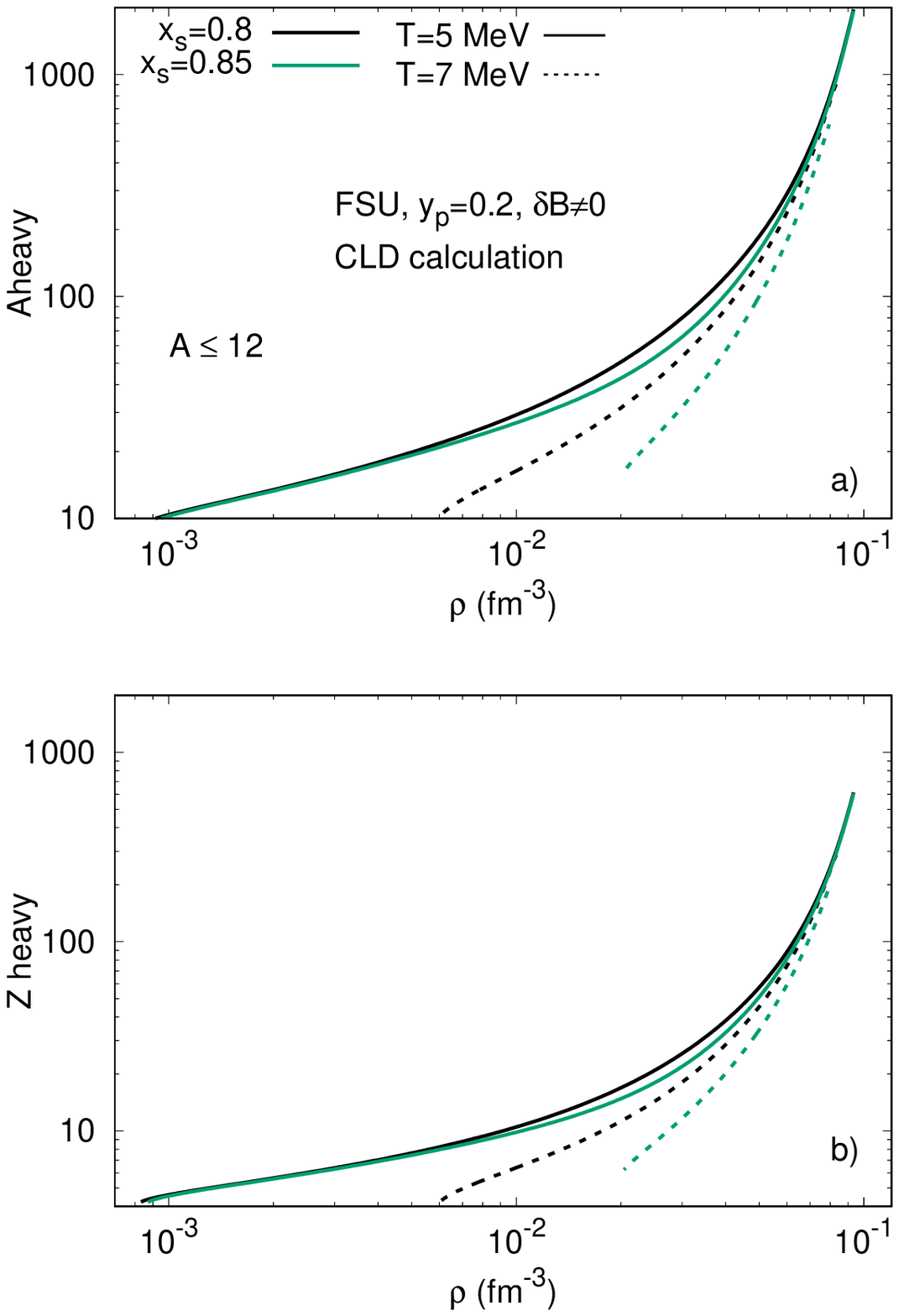} 
  \end{tabular}
  \caption{Number of nucleons, $A$, (top), and charge content, $Z$, (bottom), as a function of the density, in a CLD calculation with $\delta B$ and $A \leq 12$.  The results are for the FSU model, and a proton fraction of $y_p=0.2$, for $x_s=0.8$ (black), and $x_s=0.85$ (green), with $T=5$ (solid) and 7 (dashed) MeV. } 
\label{fig11}
\end{figure}

\begin{figure*}
 \begin{tabular}{c}
 \includegraphics[width=0.99\textwidth]{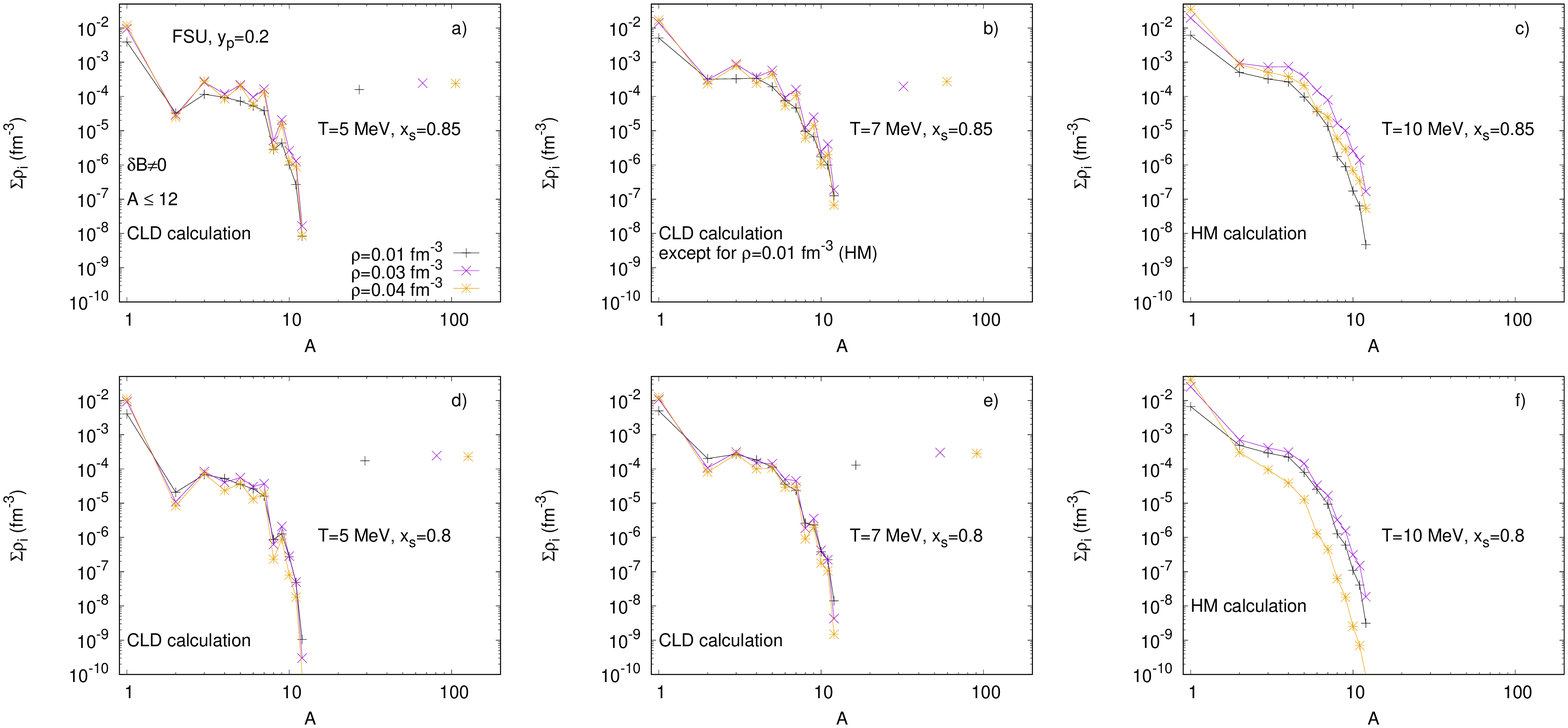} 
 \end{tabular}
  \caption{Total density of light clusters summed over $N$, i.e. $\rho(A)=\sum_{N=1}^{8}\rho_N(A,N)$,  as a function of $A$ for a fixed total density of $\rho=0.01$ (black), $0.03$ (magenta),  and $0.04$ (orange) in a CLD calculation with $\delta B$. The results are for FSU, $T=5$ (left), and 7 (middle) MeV, with $y_p=0.2$, and $x_s=0.85$ (top) and 0.8 (bottom). Note that for $A=1$, we show $(\rho_{n_2}+\rho_{p_2})(1-f)$. We also show $A_{heavy}$, and the correspondent density, $\rho_{\rm heavy}$, given by eq. (\ref{rhoCLD}). The right panels show results for a homogeneous matter calculation with light clusters for $T=10$ MeV.}
\label{fig12}
\end{figure*}

In Figs.~\ref{fig7} and \ref{fig8}, we show the mass fractions of the free nucleons, of the light clusters, taking also separately the ``exotic" and ``classical" contributions, and of the heavy cluster, for $T=5$ MeV, and the cluster-meson coupling $x_s=0.8$, considering different calculations with (CLD+cl) and without (HM+cl) the heavy cluster. 

In Fig.~\ref{fig7}, we compare a calculation without the heavy cluster (dashed), as discussed in the previous sections, with the new CLD calculation with light clusters (solid), for a fixed proton fraction of 0.2. We see that the dissolution densities of the light clusters in the CLD calculation happens after the HM calculation, and, because of the heavy cluster, the mass fractions of the light clusters are smaller in the CLD calculation. These two calculations were done taking into account the binding energy shift contribution, $\delta B_i$, in the total binding energies of the light clusters.

Let us now discuss the effect of including this term, $\delta B_i$, in the CLD+cl calculation. In Fig.~\ref{fig8}, we compare the CLD+cl calculation with (solid) and without (dashed) the inclusion of the binding energy shift, for a fixed proton fraction of 0.2 (top) and 0.41 (dashed). This term has no effect on the densities at the onset of the heavy cluster, but  an important finite effect is seen close to  the melting density for the $y_p=0.2$ calculation. The light clusters  appear in smaller abundancies and dissolve at much lower densities, and the heavy clusters are 
more massive, when taking this term into account. However, the extra binding energy term has a negligible contribution in the $y_p=0.41$ calculation. This reflects the fact
 that, for nuclear matter with a small asymmetry, the background gas density of
 nucleons is small, in the range of values for which the extra binding
 term does not play a role. 

Let us now discuss the effect on the  heavy cluster distribution  of simultaneously including the light clusters and heavy cluster in the minimization of the free energy. In Figs. \ref{fig9} and  Fig. \ref{fig10}, we show  for a fixed proton fraction of 0.2 and 0.41 and several temperatures, the total fraction of clusters, light and heavy, in a CLD both with and without light clusters. For the cluster-meson couplings, we are taking respectively $x_s=0.8$  and $0.85$,  and, in all calculations, $x_v=1$, with $g_{s_i}=x_s g_s A_i$ and $g_{v_i}=x_v g_v A_i$. The  contribution  $\delta B$ is always included in the definition of the binding energy of the light clusters in both calculations. 
  
 The dash-dotted lines in Fig.~\ref{fig9} gives the
 results of a calculation without light clusters, and shows that, in
 this case, the onset of the heavy cluster occurs at lower densities, the background of free nucleons is smaller, and the mass  fraction of
 nucleons in the heavy clusters is larger. We see that, taking in the
 calculation a larger number of degrees of freedom through the
 inclusion of light clusters, not only reduces the size of
 the heavy cluster, but also  increases the fraction of free nucleons in the
 background gas. It seems that if the calculation is too restrictive
 with respect to the competing degrees of freedom,
 it overestimates the role of the heavy cluster, with too many
 nucleons contributing to the cluster.

We have repeated the same calculation taking $x_s=0.85$, see
Fig. \ref{fig10}. Most of the conclusions are the same as the ones drawn
for Fig. \ref{fig9}, however, there are also some differences that
are worth being discussed. A larger $x_s$ favors larger fractions of
light clusters, and a smaller role played by the heavy cluster, and it also decreases the background gas of free nucleons. It is even seen that  for the larger proton fraction, $y_p=0.41$, and $T=5$ MeV, the light clusters compete with the heavy cluster close to the transition to homogeneous matter. For this proton fraction, the heavy cluster has melted already for $T=7$ MeV.
Comparing with the calculation without light clusters (dash-dotted lines), we see how sensitive is the distribution of matter between the heavy and the light clusters for the parameter $x_s=0.85$.

The role of $x_s$ is more clearly seen in Fig.~\ref{fig11}, where the number of nucleons $A_{heavy}$ (top), as well as the  charge content, $Z_{heavy}$ (bottom), is plotted as a function of density for $x_s=0.8$ (magenta) and 0.85 (green). We consider different temperatures, $T=5$ (solid) and 7 (dashed) MeV, and a fixed proton fraction of 0.2. The  main conclusion is that the number of nucleons in the heavy
cluster at a given density decreases with increasing temperature. For the lowest temperature
 considered, $T=5$ MeV, the smallest configuration has  $A_{heavy} \gtrsim 10$,
 which is compatible with taking light clusters with $A\le 12$, and avoiding
 double counting.  Comparing the results with  $x_s=0.8$ and 0.85, we
 confirm the discussion above: for $x_s=0.85$, the range of
  densities where the heavy cluster exists is smaller, and the number of
  nucleons in the clusters is also smaller. A larger value of $x_s$ favors the
  appearance of light clusters at the expense of the nucleon content
  of the heavy cluster. A similar discussion is valid for
  the charge.

\begin{figure*}[!htbp]
 \begin{tabular}{c}
\includegraphics[width=0.99\textwidth]{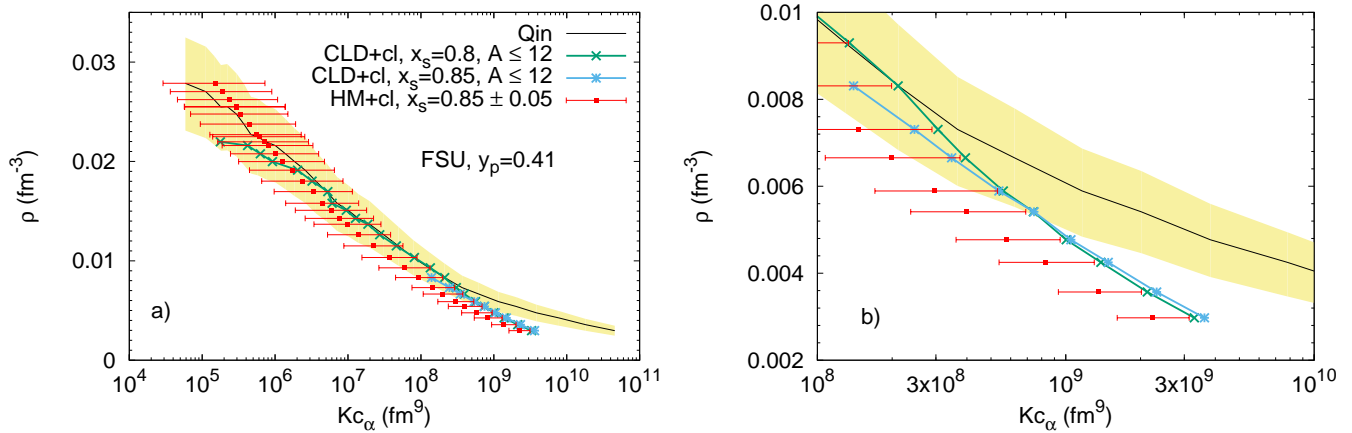} 
 \end{tabular}
  \caption{Chemical equilibrium constants of $\alpha$ for FSU, and $y_p=0.41$, and for the universal $g_{s_j}$ fitting with $g_{s_j} =(0.85\pm 0.05)A_j g_s$, considering a calculation with only the classical light clusters (red), and comparing with the CLD calculation with light clusters, taking $A \leq 12$, with the $x_s=0.8$ (green) and $x_s=0.85$ (cyan). The experimental results of Qin et al. \cite{qin12} ( solid black line and yellow uncertainty region) are also shown. The right panel shows in more detail the lowest density points of our calculations. }
 \label{fig13}
\end{figure*}

Fig. \ref{fig12} shows the total density of clusters, heavy and light,
as a function of $A$ for fixed total densities obtained with $x_s=0.85$
(top panels) and 0.8 (bottom panels). For the light clusters, the
total densities are summed over the number of neutrons, $N$,
i.e. $\rho(A)=\sum_{N=1}^{8}\rho_N(A,N)$. 

The results are for $y_p=0.2$ and $T=5$ MeV (left), and 7 MeV (middle). For $A=1$, we show the total density of protons and neutrons in phase 2, $(\rho_{n_2}+\rho_{p_2})(1-f)$, i.e. the density of the free protons and neutrons. For the heavy cluster, the correspondent density,
$\rho_{heavy}$, given by eq. (\ref{rhoCLD}), is represented.  This is calculated in the CLD approximation. In the right panels, the results shown are for a homogeneous matter (HM) calculation with light clusters with $T=10$ MeV.
We only show results obtained with the extra binding energy
term. In fact, the effect of the $\delta
  B$ term is only important for the largest density represented and
  lowest temperature. In this case, it decreases the light cluster fractions and slightly increases
  the mass number of the heavy cluster as discussed before. At larger temperatures, the extra binding of the heavier clusters does not play a big role anymore.

For the  smaller proton fraction considered, 
 the contribution of the ``exotic" clusters is more important. This can be understood from the fact  
that the most unstable light clusters are extremely neutron rich, and thus favored at low proton fractions; at the lowest density represented, the deuteron
plays an important role but for the other densities, 
heavier clusters with $A=3-5$ have larger abundancies.

It is also interesting to discuss the results obtained for $T=10$
MeV. At this temperature, the heavy cluster has already melted, and
only light clusters are present. The deuteron is the most abundant
cluster, and the larger  the  density, the larger the abundance for both values of $x_s$. The
effect of $x_s$ is clearly reflected on the results obtained for the
other clusters: a larger value of $x_s$ favors the appearance of light
clusters. For $\rho=0.04$  fm$^{-3}$, the differences between both
values are quite dramatic. Also for $T=5$ and 7 MeV, the abundance of the
light clusters is always smaller for  $x_s=0.8$.

In Fig. \ref{fig13}, the chemical equilibrium constant for the
 $\alpha$ cluster obtained in a calculation including
a single heavy cluster and light clusters with $A\le12$ is plotted as a
function of the density for $x_s=0.8$ (green line) and $x_s=0.85$
(cyan line). The figure also contains the  experimental results
of Qin et al. \cite{qin12} (solid black line and yellow uncertainty region), and results from a
calculation considering only  light clusters with $A\le4$, and
$g_{s_j} =(0.85\pm 0.05)A_j g_s$ (red marks with arrow-bars). We
recall that the different data points of Ref. \cite{qin12}  in this
plot corresponds to different temperatures. The densities below
$\rho=0.01$ fm$^{-3}$ correspond to temperatures in the range $5>T>7$ MeV.  These low-density points are shown in more detail in the right panel of Fig. \ref{fig13}. There is a discrepancy at low densities where the medium modifications are small. The
presence of the heavy cluster  shifts  the equilibrium constant at
a given density to larger values. This can be achieved either if the
$\alpha$-densities are larger, or  if the background gas is less dense. In
fact, looking at Fig. \ref{fig10}, it can be concluded that both
effects are present: the $\alpha$-density increases and the background gas density decreases.
As discussed before, a larger value of $x_s=0.85$ melts the heavy
cluster at smaller densities, and, therefore, in this case, we cannot go
above $\rho=0.01$  fm$^{-3}$.
The CLD calculation including the heavy cluster is clearly more realistic 
for the description of stellar matter. Still, the most meaningful comparison with the experimental 
data of Qin et al. \cite{qin12} is probably given by the red symbols. Indeed, in the experimental conditions of the HIC (radial flow of light particles from a central dense participant zone) the number of particles and the time scale of the reaction are certainly not enough for the realization of a thermodynamic equilibrium between light particles and heavy clusters.

\section{Conclusions}\label{sec:Conclusions}

In the present study, we have addressed the problem of the description
of warm non-homogeneous matter at subsaturation densities as the ones occurring in core-collapse
supernova or neutron star mergers.  The formalism was developed in the
framework of relativistic mean-field models.  We have completed
a previous work \cite{pais18}, by investigating the
inclusion of all light clusters with $A\le 12$, and the combined inclusion
of light clusters and one heavy cluster within the CLD approach.
For the cluster-meson couplings, we considered  the approach proposed in
\cite{pais18} for all mesons, where, in a self-consistent way, the
background nucleon gas affects the binding energy of the light
clusters. 

Including the light clusters with a larger mass number had a visible
effect in the particle distribution maximum, with a decrease of the classical light clusters. However, no noticeable effect is identified at the cluster onset or cluster
melting. The relative effect of the ``exotic" light clusters, i.e. the ones with $4<A\le
12$, is more important for the lower temperatures and smaller proton
fractions. It was also shown that generally the isotopes with smaller
$Z$ are the most abundant. An exception occurs for the lower
temperatures close to the peak of the distributions where  $Z=2$
isotopes, or even  $Z=3,4$ isotopes, for a large value of $y_p$, may
become more abundant.

For the application of light cluster production in heavy ion collisions, taking into account the decay schemes of the ``exotic" light clusters, we
have defined effective $^4$He and $^3$H cluster abundances that
include the corresponding cluster abundances plus the  contribution
of the ``exotic" clusters that decay into these ones. We could show that,
while at low temperatures,  the classical light clusters alone could
simulate the contribution of the light clusters, for the higher
temperatures or higher proton fractions, this is not anymore the
case: close to the peak of the cluster distributions, there is a clear
reduction on the proton and neutron background gas, and increase of the 
$^4$He and $^3$H cluster abundances. The non-equivalence is also
reflected on the equilibrium constants, especially at the higher densities.
On the other hand, taking the set of clusters  with $A\le 12$ or
just the ones with $A\le4$ does not affect much the equilibrium
constants. This is due to the fact that, although the fraction of the
classical light clusters is smaller, also the fraction of nucleons in
the background gas is smaller. Taking effective $^4$He and $^3$H
cluster abundances, these ones may get larger than the ``classical"
light clusters distributions with a smaller fraction  of background
gas nucleons.

Taking into account the combined contribution of a heavy cluster and
the light clusters we could conclude that the presence of a heavy
cluster reduces in general the contribution of light clusters, but also
shifts the melting density to higher densities. As a consequence,
light clusters may still occur at larger densities, as compared to the range of densities within a calculation where only light clusters are included. It was also shown that including the background gas contribution on the binding energy of the
light clusters has an important effect on the light cluster
abundances, shifting down the melting densities and decreasing the
light cluster abundances. As a consequence, the heavy clusters will be
heavier with a larger proton number, and the  light clusters
with $A>2$ will be less abundant.

\section*{ACKNOWLEDGMENTS}
This work was partly supported by the FCT (Portugal) Projects No. UID/FIS/04564/2016 and POCI-01-0145-FEDER-029912, and by PHAROS COST Action CA16214. H.P. is supported by FCT (Portugal) under Project No. SFRH/BPD/95566/2013. She is very thankful to F.G. and her group at LPC (Caen) for the kind hospitality during her stay there within a PHAROS STSM, where this work started.

\end{document}